\documentclass[prb,twocolumn,superscriptaddress,letterpaper,amssymb,amsmath]{revtex4-1}
\pdfoutput=1

\usepackage{amsmath}
\usepackage{stmaryrd}
\usepackage{color}
\usepackage{graphicx}

\begin{document}

\title{Luttinger-volume violating Fermi liquid in the pseudogap phase of the cuprate superconductors}

\author{Jia-Wei Mei} \affiliation{Institute for Advanced Study, Tsinghua
University, Beijing, 100084, P. R. China} \affiliation{Department of Physics,
Massachusetts Institute of Technology, Cambridge, Massachusetts 02139, USA}
\author{Shinji Kawasaki} \affiliation{Department of Physics, Okayama University,
Okayama 700-8530, Japan}
\author{Guo-Qing Zheng} \affiliation{Department of Physics, Okayama University,
Okayama 700-8530, Japan} \affiliation{Institute of Physics and Beijing National
Laboratory for Condensed Matter Physics, Chinese Academy of Sciences, Beijing
100190, China}
\author{Zheng-Yu Weng} \affiliation{Institute for Advanced Study, Tsinghua
University, Beijing, 100084, P. R. China}
\author{Xiao-Gang Wen} \affiliation{Department of Physics, Massachusetts
Institute of Technology, Cambridge, Massachusetts 02139, USA}
\affiliation{Institute for Advanced Study, Tsinghua University, Beijing,
100084, P. R. China}

\date{\today}

\begin{abstract}
Based on the NMR measurements on Bi$_2$Sr$_{2-x}$La$_x$CuO$_{6+\delta}$ (La-Bi2201) in strong magnetic fields, we identify the non-superconducting pseudogap phase in the cuprates as a Luttinger-volume violating Fermi liquid (LvvFL). This state is a zero temperature quantum liquid that does not break translational symmetry, and yet, the Fermi surface encloses a volume smaller than the large one given by the Luttinger theorem. The particle number enclosed by the small Fermi surface in the LvvFL equals the doping level $p$, not the total electron number $n_e=1-p$. Both the phase string theory and the dopon theory are introduced to describe the LvvFL. For the dopon theory, we can obtain a semi-quantitative agreement with the NMR experiments.
\end{abstract}

\maketitle

\section{Introduction} \label{sec:intro}

When the high $T_c$ cuprate superconductor was first discovered, everyone knew what is the key issue for solving the high $T_c$ problem: finding a right pairing mechanism that gives rise to high $T_c$. This point of view is based on the strong belief that the normal metallic state above $T_c$ is described by Fermi liquid theory\cite{[See the classic book ]Abrikosov1963} and the superconductivity is caused by a pairing instability of the Fermi liquid. After 25 years of intensive research on high $T_c$ superconductors, a lot of progress has been made. One sign of the progress is that it is no longer clear, or people no longer agree on, what is the key issue for solving the high $T_c$ problem. This is because for the underdoped cuprates, it is not clear whether the normal metallic state above $T_c$ is described by Fermi liquid theory. It is also not clear whether the superconductivity is caused by the pairing instability of the Fermi liquid, since we may not have a Fermi liquid above $T_c$.

In this paper, we take a point of view that the key issue for solving the high $T_c$ problem is not to find a pairing mechanism, but to first understand the normal metallic state above $T_c$. Only after well understanding the normal metallic state above $T_c$, can we start to address the secondary question: what causes the instability of the normal metallic state above $T_c$ towards superconducting state. For example, if the normal state did not even have quasiparticles, then it would be impossible to understand the high $T_c$ superconductivity via the paring of quasiparticles.

The standard Landau Fermi liquid has two key features:\cite{Abrikosov1963} (i) it has well-defined low energy quasiparticles; (ii) the particle number enclosed by the Fermi surface equals the total electron number $n_e$ satisfying the Luttinger's theorem.\cite{Luttinger1960,Oshikawa2000} For overdoped cuprates, it appears that the normal metallic state above $T_c$ is a Fermi liquid. It has well-defined low energy quasiparticles and larger Fermi surface enclosing the total number of the electrons $n_e$.\cite{[For references see] Damascelli2003,Plate2005,Vignolle2008} So the key issue for the overdoped cuprates is to understand the pairing mechanism of the superconducting instability.  In underdoped samples, the normal metallic state is a pseudogap (PG) state. The quasiparticles with sharp peaks are observed near the nodal region in the ARPES experiments above $T_c$.\cite{Meng2009,Meng2009a} The quasiparticles are also well-defined enough to have the quantum oscillations in strong magnetic fields.\cite{Doiron-Leyraud2007} However, the PG state in the underdoped cuprates do not have large Fermi surface enclosing the total number of electrons $n_e$. Instead it has small Fermi surfaces observed in the quantum oscillation measurements. \cite{Doiron-Leyraud2007} Low energy quasiparticles are also found only around the nodal region in the ARPES.\cite{Meng2009} The normal state above $T_c$ for the underdoped samples is no longer the standard Fermi liquid. It has well-defined low energy excitations, but small Fermi surfaces with the volume violating the Luttinger's theorem. So the key issue is to understand such a new normal metallic state, which is called the PG metallic state (PGMS) in this paper.

In the experiment, the robust $d$-wave superconductivity conceals the PGMS at low temperatures. The experiments at finite temperatures near $T_c$ contain too many excitations, that also conceal the low energy properties of PGMS, such as if Fermi surface and/or quasiparticles exist or not. To sharpen the issue on PGMS in the experiments, we achieve the non-superconducting normal state at zero temperature by applying high magnetic fields (up to 44T) on the low $T_c$ cuprate $\text{Bi}_2\text{Sr}_{2-x}\text{La}_x\text{Cu}\text{O}_{6+\delta}$ (La-Bi2201).\cite{Zheng2005,Kawasaki2010} All the measurements are performed in the magnetic fields along the $c$-axis. In T=0K limit, the PGMS  has a reduced finite DOS and appears to violate the Luttinger's theorem. Many efforts have been made trying to  rescue the Luttinger's theorem in the PGMS. Most among them introduce long range orders breaking the translational symmetry to cut the large Fermi surface into small pieces. However, no translational symmetry breaking is found in the PGMS in our measurements.

Based on the NMR measurements, in this paper we propose that the PGMS in the underdoped high-$T_c$ cuprates is a metal with properties as follows\\
(A) The PGMS has few low energy excitations, as reflected by a low but finite effective ``density of states'' (DOS) in the $T=0$ limit.\\
(B) The PGMS is a quantum liquid at zero temperature.\\
(C) The PGMS does not break the translation symmetry.\\
(D) The low energy excitations associated with small effective DOS are described by quasiparticles with small Fermi surfaces that are not large enough to enclose large Fermi volume.\\
Such a new quantum metallic state of electrons is not a standard Landau Fermi liquid. It has few low energy quasiparticle excitations in T=0 limit, and small Fermi surfaces with the volume violating the Luttinger's theorem. We call the PGMS a Luttinger-volume violating Fermi liquid (LvvFL) in this paper.

To give a theoretical description of the LvvFL state, we start from the Mott insulator at half-filling (that does not break the translation symmetry) and describe the underdoped cuprate as a doped Mott insulator.\cite{Lee2006} If the dopant holes/electrons do not interact too strongly with underlying spins, the dopant holes/electrons may form small Fermi pockets whose volume is equal to the doping density. Different from the Fermi liquid, the Mottness leads to the sparse fermionic signs:\cite{Wu2008} the Mottness at half-filling is robust and doesn't carry the fermionic signs; only the dopant holes in the cuprates carry the fermionic statistics and these fermions occupy small Fermi pockets resulting in small DOS in the NMR experiments.

In this paper we employ two different theories (the phase string theory\cite{Weng2007, Weng2011, Weng2011a} and the dopon theory\cite{Ribeiro2005, Ribeiro2006}) as two examples to describe the LvvFL in the PGMS. Both theories describe the LvvFL in terms of two different components: a background resonating valence bond (RVB) state (robust Mottness at half-filling) and a Fermi liquid with small Fermi pockets (the dopant holes). In this paper, we mainly focus on the dopon theory and describes the LvvFL in terms of two different fermions: ``spinons'' are charge neutral spin-1/2 excitations of the background RVB; ``dopons'' have the charge $+e$ and spin-1/2 degrees of freedom as the same as the dopant holes. At the mean field level, the spinons and dopons do not hybridize in the LvvFL. The dopon sector has the hole pockets strongly renormalized by the background antiferromagnetic spin background similar to the numerical results in Ref.  \onlinecite{Dagotto1994}. At the mean field level, the dopon theory gives a semi-quantitative result in a good agreement with the NMR experiment.

\section{Experimental aspects}
\subsection{The PGMS has few low energy excitations}\label{sec:ldos}

The doping regime for the PG state in La-Bi2201 is $0.10<p<0.21$ with the corresponding La concentration $0.2<x<0.95$.\cite{Zheng2005,Kawasaki2010} A large PG in the single particle spectrum was also confirmed in the ARPES\cite{Meng2009,Meng2009a} and STM/STS\cite{Sugimoto2006} experiments. The resistivity in the PG regime was measured under a high pulsed magnetic field (60T) when the superconductivity was completely suppressed.\cite{Ono2000} A crossover from metal to an ``insulating'' state was found near $p_c=1/8$ ($x=0.65$): for the larger doping $p>p_c$, the in-plane resistivity remains finite in $T=0$ limit; such a metallic behavior changes into $\log{(1/T)}$ type insulating behavior in the lower doping regime at $p<p_c$.\cite{Ono2000} Here the insulating behavior is not due to a full gap opening in the DOS, but some peculiar charge localization. So we will use the term ``PGMS'' for the whole PG regime.

The optimal La-Bi2201 ($p=0.16$ with $x=0.4$ and $T_c=32 \text{K}$) has a PG temperature $T^*\sim 150\text{K}$. Below $T^*$, the Knight shift (proportional to the DOS) starts to decrease with decreasing temperature, see Fig. \ref{fig:chi}(c). Such a PG-type decreasing in DOS is independent of the magnetic field before entering the superconducting state ($T>T_c(H)$). When the magnetic field increases beyond $H_{c2}\simeq26\text{T}$, the superconductivity is killed completely and the PGMS is realized in $T=0$ limit. A finite DOS in PGMS declares the existence of low energy excitations at zero temperature.

The hyperfine coupling constant in the optimal sample is $A_{hf}=194\text{KOe}/\mu_B$.\footnote{S. Kawasaki, unpublished.} In $T=0$ limit, the Knight shift in PGMS (with $H>H_{c2}$) is $K_c=1.36\%$ and the orbital contribution is $K_{\text{orb}}=1.21\%$.\cite{Kawasaki2010} Therefore, the residual finite DOS of the electrons is given by 
\begin{eqnarray}
  \frac{dn}{d\epsilon}=\frac{K_c-K_\text{orb}}{A_{hf}a^2\mu_B^2}
  \label{eq:dos}
\end{eqnarray}
Thus we find $\frac{dn}{d\epsilon}=5.78\times10^{26}\text{erg}^{-1}\text{cm}^{-2}=5.78\times10^{37}\text{J}^{-1}\text{m}^{-2}$ by using the lattice constant $a=3.8\times 10^{-8}\text{cm}$ and $\mu_{B}=9.27\times 10^{-21} \text{emu}$.  Based on the Fermi pocket scenario in this paper, we can also estimate the DOS by 
\begin{eqnarray}
  {\frac{dn}{d\epsilon}}={\frac{1}{2\pi ^{2}}}{\frac{L_{k}}{\hbar v_{F}}}
  \label{eq:esdos}
\end{eqnarray}
where the Fermi velocity $v_{F}$ can be obtained in ARPES\cite{Meng2009a} as $v_{F}=1.75\text{eV}\cdot \text{\r{A}}$ and the Fermi surface length has the approximate value $L_{k}=2\pi \sqrt{2\pi n_{fp}p}/a$ with $n_{fp}$ denoting the total number of the Fermi pockets in the Brillouin zone. In the dopon theory (see Sec. \ref{sec:dopon}), $n_{fp}=4$ and one can estimate $\frac{dn}{d\epsilon }=5.9\times 10^{37} \text{J}^{-1}\text{m}^{-2}$, which is very close to the above experimental value.

\subsection{The PGMS is a quantum liquid}
\begin{figure}
  \includegraphics[width=\columnwidth]{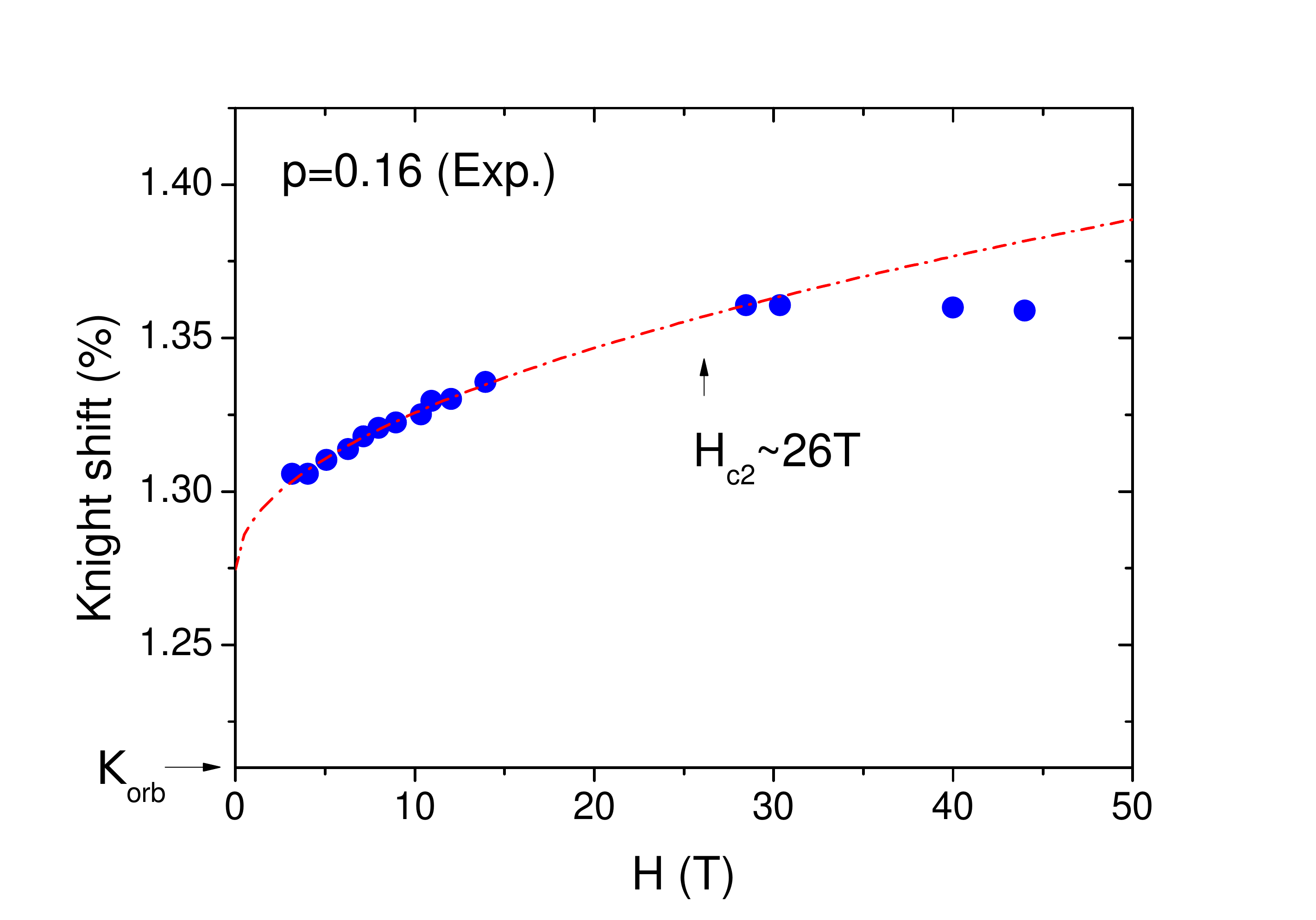}\\
  \caption{[Color online] $H$-dependence of the Knight shift for the optimal La-Bi2201. The dash
  line is the fitting curve $K_c=1.27+0.016\sqrt{H}$ at $T=4.2~\text{K}$. The
  square-root behavior breaks down when $H>H_{c2}$. The orbit contribution to the Knight shift is $K_\text{orb}=1.21\%$. The DOS is finite even at $T=0$ and $H=0$ due to impurity in the $d$-wave superconductor. }
\label{fig:square_root}
\end{figure}

Firstly, we can rule out the ``$d$-wave superconductor+thermal fluctuation" (dSC+TF) scenario. In dSC+TF scenario, the $d$-wave superconducting gap remains finite in the PG state and the thermal fluctuations release the DOS near the nodal lines on the Fermi surface. In this picture, the PG phase is a disorder dSC state and a nodal metal. It is a thermal liquid and the finite DOS is a thermal effect. Due to the Volovik effect,\cite{Volovik1993} the DOS has a square-root behavior ${ \frac{dn}{d\epsilon }}\propto \sqrt{H}$ when the $d$-wave superconducting gap remains finite as the magnetic field is applied.  For the optimally doped La-Bi2201, the Volovik effect has been indeed observed in the Knight shift measurement when $H<H_{c2}$ in $T=0$ limit.  However, the DOS becomes independent of the magnetic field for $H>H_{c2}$, see Fig. \ref{fig:square_root}, where the Volovik square-root behavior breaks down. So the dSC gap in the PG state should vanish and the dSC+TF is ruled out based on our NMR data. As shown in Fig. \ref{fig:square_root}, we can determine the upper critical field in the optimal La-Bi2201, $ H_{c2}\simeq 26\text{T}$. Compared with the other experiments, this value is very close to the melting field $H_{m}(T)\simeq 25\text{T}$ and much smaller than the ``upper critical field'' $H_{c2}$$ \simeq 50\text{T}$ determined in the Nernst measurements.\cite{Wang2006, Li2010}

\begin{figure}
  \includegraphics[width=\columnwidth]{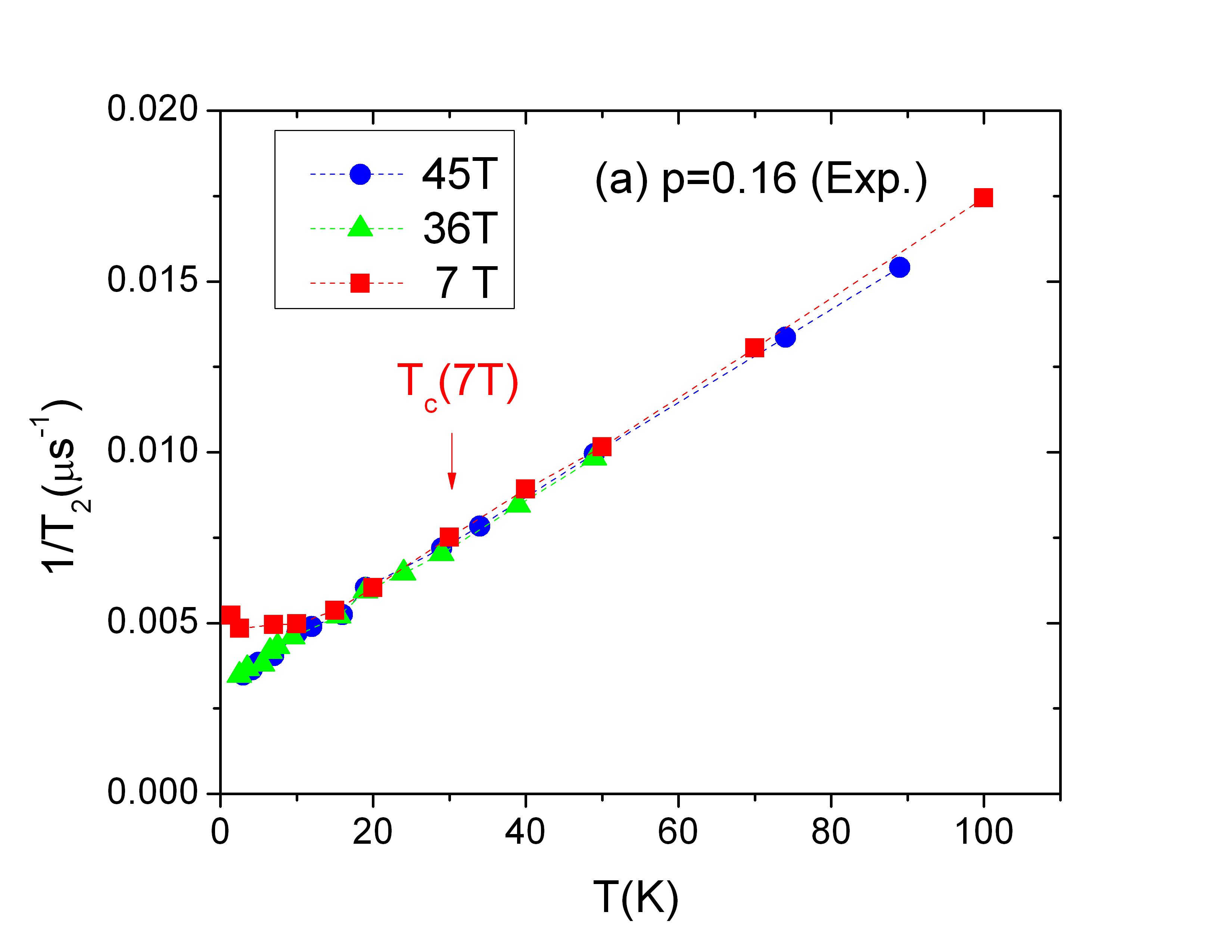}\\
  \includegraphics[width=\columnwidth]{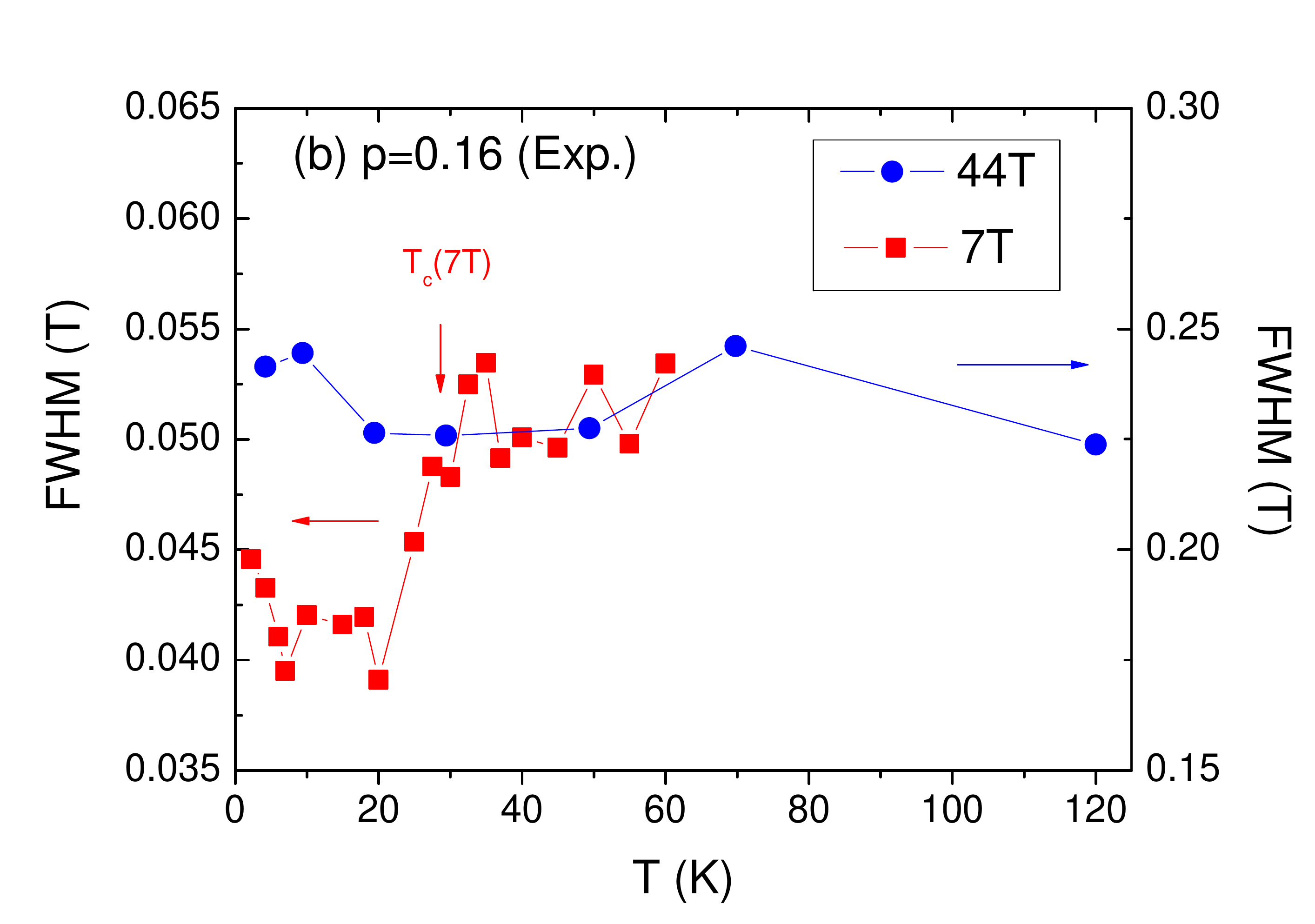}\\
  \caption{[Color online]  (a) $T$-dependent $1/T_2$ on $^{63}\text{Cu}$ under
  different magnetic fields for the optimal La-Bi2201;  (b) $T$-dependent FWHM
  for the optimal La-Bi2201 under the magnetic field 7T under $H_{c2}$ and 44T
  above $H_{c2}$. A clear SC drop is shown in FWHM under 7T.}\label{fig:T_2}
\end{figure}
Our second concern is the vortex issue in the PG state. In the superconducting state, the applied magnetic field will create magnetic vortices. These magnetic vortices lead to the Volovik square-root behavior in the DOS when $H<H_{c2}$, as shown in Fig. \ref{fig:square_root}. We are concerned about whether vortices, if any, form a thermal or quantum liquid in PG state. In Ref. \onlinecite{Recchia1997}, the vortex thermal dynamics with slow thermal motion was observed in the transverse nuclear magnetic relaxation rate $1/T_{2}$ in YBa$_{2}$Cu$_{3}$O$_{7}$. The T-dependent $ 1/T_{2}$ under different magnetic fields is also measured on $^{63}$Cu in the optimal La-Bi2201. Above $H_{c2}$, the PG-type decreasing behavior is also found in $1/T_{2}$ and independent of the magnetic fields as shown in Fig. \ref{fig:T_2} (a), consistent with the T-independent PG-type behavior for the Knight shift in Fig. \ref{fig:chi} (c). With decreasing the magnetic field below $H_{c2}$, there is some signal in $1/T_{2}$ associated with the onset of SC at low temperatures, showing a clear enhancement in the SC state, possibly due to the magnetic vortices. The low temperature SC behavior is also observed in the full width at half maximum (FWHM) of the intensity on $^{63}$Cu . It has a clear drop below $T_{c}(7\text{T})$ at 7T as shown in Fig.\ref{fig:T_2} (b). At 44T, there is no such a superconducting drop in FWHM indicating that the superconductivity is killed completely under strong magnetic fields. This is also an evidence indicating vanished SC gap in the PGMS. So the experiment does not support a thermal liquid of vortices in the PGMS near $T=0$. At $H>H_{c2}$, the vortices, if any, should form a quantum liquid.

The last concern is how we should view such a metallic state in the presence of magnetic field. We would like to argue that, due to impurities, such a state is very different from the quantum Hall states. The optimal La-Bi2201 has a resistivity $\rho=0.12~\text{m}\Omega\text{cm}$ when the superconductivity is killed at 0K. Then the mean free path is around $l={ \frac{hl_c}{e^2\rho}}/4\tilde{k}_F=13 a$, $a$ and $l_c$ are the in-plane and c-axis lattice constant, respectively; the Fermi vector is $\tilde{k}_F=\sqrt{ \pi p/2}$. At 30T magnetic field, the magnetic length $l_B=\sqrt{2\pi\hbar/eB }=31 a$. The mean distance between dopant holes is $a\sqrt{1/p}=2.5a$. We see that and $a\sqrt{1/p}<l<l_B$. The mean free path is long enough to have a well-defined wave vector of the electrons and also short enough to suppress the orbit effect of the magnetic field. So we should view PGMS in the presence of magnetic field as a dirty metal. There is no quantum Hall effect since $l<l_B$.

\subsection{The PGMS has the small Fermi surface}

Below we will also rule out the large Fermi surface scenario. Within the large Fermi surface scenario, the PG behavior of the Knight shift, i.e., the decreasing DOS with decreasing the temperature, should be caused by the Fermi velocity increasing. However, such an increasing of the Fermi velocity at low temperature in PGMS is not observed in the optimal La-Bi2201. The Fermi velocity $v_{F}=1.75\text{eV}\cdot \text{\AA}$ can be obtained from ARPES\cite{Meng2009a} at high temperature. In $T=0$ limit, the gap slope is $v_{\Delta }=0.028~\text{eV}\cdot \text{\AA}$ extracted from the specific heat.\cite {Wang2011} The ratio $v_{F}/v_{\Delta }$ can be obtained from the thermal conductivity measurement,\cite{Ando2004} $v_{F}/v_{\Delta }=33$. Thus the Fermi velocity may be estimated as $v_{F}=0.93~\text{eV}\cdot \text{\AA}$ at low temperature. No Fermi velocity increasing is found with decreasing the temperature to support a PG behavior in the large Fermi surface scenario.

In Sec. \ref{sec:ldos}, we estimate the finite residual DOS in PGMS at $T=0$K and find a very good agreement between the experimental and theoretical data based on the Fermi pockets. In other words, the small Fermi pockets are more natural to account for the NMR experiment in the PGMS. \footnote{The Fermi arc is also possible, however, in this paper we use the Fermi pocket scenario.}

\subsection{The PGMS does not break translational symmetry}
\begin{figure}
  \includegraphics[width=\columnwidth]{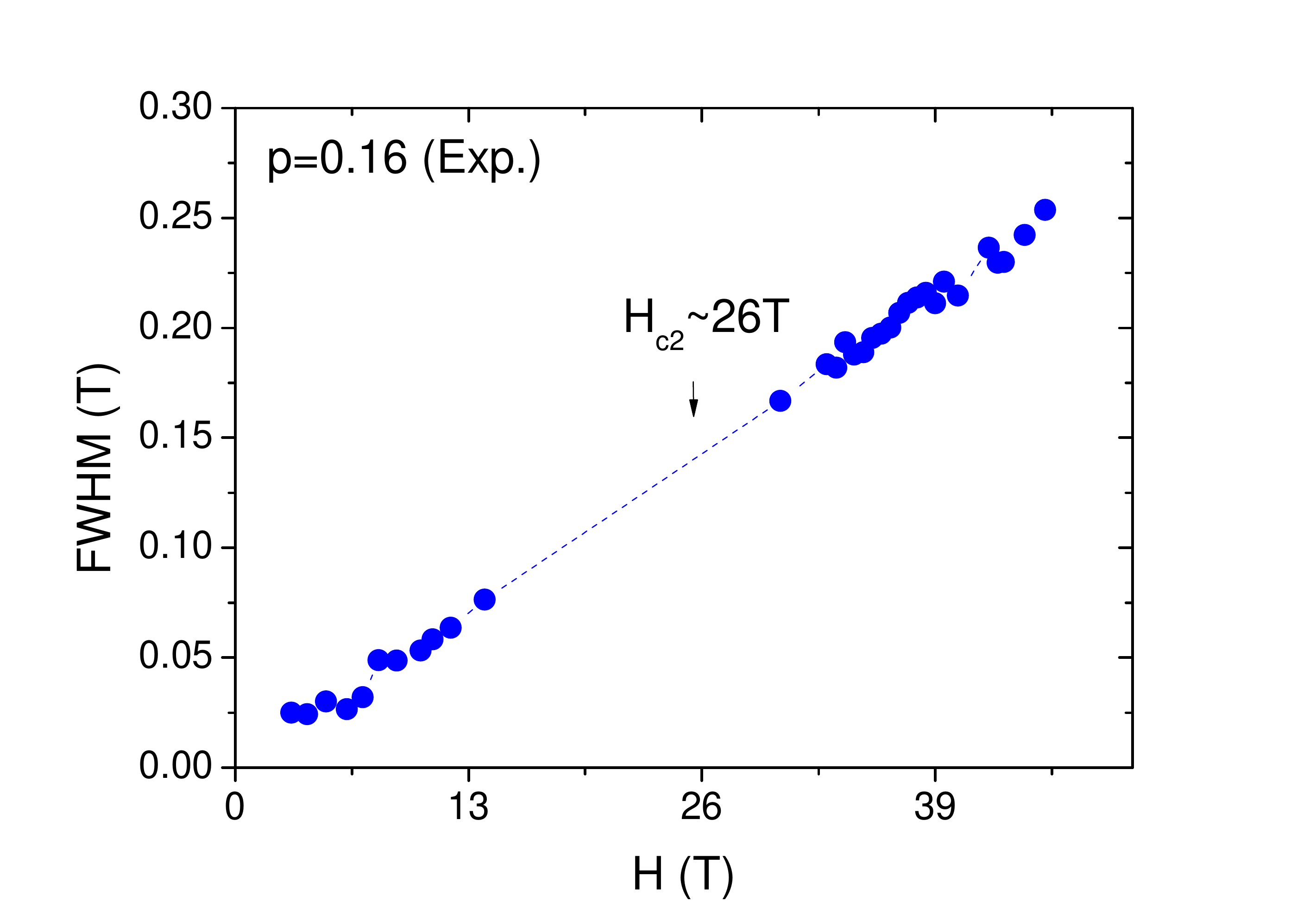}\\
  \caption{[Color online] $H$-dependent FWHM in optimal La-Bi2201 at T=4.2K.}\label{fig:fwhm}
\end{figure}

If there is a translational symmetry breaking, the long range orders truncate the large Fermi surface into small pieces. The NMR/NQR experiment \cite{Zheng2005,Kawasaki2010} is desired to investigate whether the PGMS has the translational symmetry or not. When the translational symmetry breaks, there will be an internal field around Cu local environment. In the heavily underdoped samples ($x=0.90$ and 0.80), internal magnetic fields at Cu site are very large, $H_\text{int}=7.2\text{T}$.\cite{Kawasaki2010}

In the optimal La-Bi2201, we find that there is no internal magnetic field in the PGMS associated with the translational symmetry breaking. Fig. \ref {fig:fwhm} shows the NMR FWHM as a function of the magnetic field. The FWHM increases linearly with increasing $H$, indicating that the FWHM is only due to the electronic inhomogeneity that gives rise to a distribution of the spin susceptibility (Knight shift), which is a typical behavior seen in the cuprates. Namely, the good linearity indicates that there is no additional broadening due to the internal magnetic field associated with the ordered moment. We can estimate an upper limit for the ordered moment, if any. At $H$=30 T, the FWHM is 0.16 T. With reasonable assumption that we can detect an additional broadening of 1/10 FWHM, and with the knowledge of the hyperfine coupling constant, $A_{hf}=194\text{KOe}$, we conclude that the magnetic field-induced moment, if any, should be less than 1 m$\mu_B$. Likewise, we can conclude that there is no charge order which would give rise to a quadrupole splitting for the first satellites of the spectrum since the distance between the two satellites is proportional to the charge state of Cu. With a conservative assumption that we can detect a splitting that is 1/2 FWHM, then we can rule out a charge order that gives rise to a splitting of $\gamma \times0.08$ T $=0.9$ MHz. This is equivalent to saying that there is no field-induced charge disproportion of 0.0075 hole or 1\% of the Cu charge state at zero magnetic field, since one Cu-3d hole would give rise to 117 MHz of the nuclear quadrupole frequency\cite{Zheng1995}.

Although the conventional translational symmetry breaking (SDW or CDW) can be ruled out by our NMR measurement, the staggered flux issue\cite{Hsu1991} remains unsolved. The internal field on the Cu sites is always zero regardless of the existence of the staggered flux in the PG state. In the recent NQR measurement in YBa$_2$Cu$_4$O$_8$, the staggered flux state is ruled out in the PGMS.\cite{Strassle2011}

\subsection{Experimental conclusions}
The PG behavior of the optimal La-Bi2201 is confirmed in the Knight shift, $1/T_{1}T$ and $1/T_{2}$. The key feature is the loss of DOS in the PG state with decreasing the temperature. However, the loss of DOS is not complete. When the superconductivity is completely killed in the strong magnetic fields, the PGMS still has few low energy excitations even in $T=0$ limit. The Volovik square-root behavior breaks down and the dSC gap vanishes. So the PGMS is not a disorder dSC state. We measure $1/T_{2}$ to address the issue whether the vortices in the PGMS, if any, form a thermal or quantum liquid and conclude that the vortices should be quantum objects if they do exist in the PGMS.

Furthermore, there is no Fermi velocity increasing with decreasing the temperature such that a large Fermi surface with Fermi velocity increasing is very unlikely to explain the PG-type decreasing in the DOS. Small Fermi pockets are thus very natural to account for the NMR data. Since the PGMS has the translational symmetry, the small Fermi pockets should not be due to the truncation of the large Fermi surface as the result of the translational symmetry breaking. The PG state above $T_c$ in the optimal La-Bi2201 has well-defined quasiparticles near the nodal region,\cite{Meng2009,Meng2009a} resembling the feature of a standard Landau Fermi liquid. However, it has small Fermi pockets without the translational symmetry breaking which violates the Luttinger's theorem explicitly. In this paper, we call the PGMS a Luttinger-volume violating Fermi liquid (LvvFL).

The PGMS is a long-standing issue in the underdoped high $T_c$ cuprates. One key point in our NMR measurements is that the PGMS is achieved in $T=0$K limit by using the static strong magnetic fields. It would be highly desirable if further measurements can be done to support or falsify the present scenario. Recently, the magnetic-field-induced charge-stripe order was found in underdoped YBa$_{2}$Cu$_{3}$O$_{y}$.\cite{Wu2011} But $T_c$ is too high for the SC phase in YBCO to be fully suppressed and the stripe order still coexists with the SC order. There is also evidence for charge-density-wave in the underdoped La-Bi2201, although none for the optimal La-Bi2201.\cite{Rosen2011} The identification of a specific phase remains elusive in these experiments. To our best knowledge, it still safe for us to conclude that the PGMS in the optimal La-Bi2201 is an LvvFL.

\section{Theoretical descriptions}
There are two main different approaches to understanding the LvvFL: one starts from the Fermi liquid at the overdoped side (Fermi-liquid approach) and the other one is from the Mott insulator at half-filling (Mott-insulator approach).

The Fermi-liquid approach tries to understand the transition from the large Fermi surface to the PG-type behavior. The large-Fermi-surface normal state with the pairing fluctuations can give an explanation of the appearance of the Fermi arc at \emph{finite} temperatures in the ARPES measurements. \cite{Engelbrecht1998,Senthil2009} In the Luttinger's original paper\cite{Luttinger1960}, the fluctuations (not only the pairing ones) may reduce the spectral weight near the Fermi surface, but cannot modify the volume of the Fermi surface. To rescue the Luttinger's theorem, long range orders that break translational symmetry are needed to cut the large Fermi surface into small pieces.\cite{Senthil2009} Yang, Rice and Zhang introduced Yang-Rice-Zhang (YRZ) ansatz for the single electron Green's function to solve such a problem.\cite{Yang2006, [For a review see ]Rice2011} The YZR Green's function satisfies the Luttinger's theorem in a different way in which the electron density is related not to the Fermi surface volume, but to the volume of singularities of logarithm of the Green's function at zero frequency.\cite{Abrikosov1963} YZR approach can result in small Fermi pockets with the quantized volume enclosing the number of the dopant holes $p$.


The Mott-insulator approach considers the effect of introducing holes into the Mott insulator state at half-filling and describe the underdoped cuprates as the doped Mott insulator. Sachdev and collaborators \cite{Qi2010,Moon2011} phenomenologically model the doped Mott insulator in the underdoped insulator as a Fermi liquid of holes moving in a background with local antiferromagnetic order. They describe the PGMS as the fractional Fermi liquid proposed by Senthil \textit{et al} in the Kondo lattice model for heavy fermions.\cite{Senthil2003,Senthil2004} Tesanovic and collaborators describe the PGMS as the quantum disordered $d$-wave superconducting state.\cite{Tesanovic2008,Tesanovic2011} They model the PGMS as the algebraic Fermi liquid in which the Fermi surface has four points just like the $d$-wave superconductor and a branch-cut with a small anomalous dimension due to the quantum fluctuations.\cite{Franz2001,Franz2002,Rantner2001}

In the following, we use another two theories, the phase string theory and the dopon theory, to describe the LvvFL identified in our NMR measurements. Similar to Refs. \onlinecite{Qi2010,Moon2011}, both theories describe the LvvFL in terms of two different components: a background resonating valence bond (RVB) state (robust Mottness at half-filling) and a Fermi liquid with small Fermi pockets (the dopant holes).

\subsection{Sparse fermionic signs and small Luttinger's volume}
For the overdoped samples, the normal metallic state in La-Bi2201 is the Fermi liquid. As the Fermi gas, the Fermi liquid has the full fermionic statistics: when two of the electrons exchange with each other, there is a negative sign in the partition function
\begin{eqnarray}
  Z_{\text{FL}}=\sum_{\{c\}}(-1)^{N_\text{ex}[c]}Z_0[c]
\end{eqnarray}
where $\{c\}$ contains all the closed loops of the spatial trajectories of the fermions and $Z_0[c]>0$ is the classical partition function on the loop $c$. The fermionic statistics is determined by the total number of electron exchanges $N_\text{ex}[c]$. The Fermi liquid has the well-defined Fermi surface with the quantized volume determined by the Luttinger's theorem\cite{Luttinger1960,Oshikawa2000}
\begin{eqnarray}
 2{\frac{V_F}{(2\pi)^2}}=n_e\mod2.
\end{eqnarray}
where $V_F$ is the Fermi volume occupied by the electrons in the $\mathbf{k}$ space.

In the doped Mott insulator, Wu \textit{et al} derived the sign structure of the $t-J$ model upon doping.\cite{Wu2008} In this case, the partition function is 
\begin{eqnarray}
  Z_{t-J}=\sum_{\{c\}}\tau_c\mathcal{Z}[c]
\end{eqnarray}
where $\mathcal{Z}[c]>0$. For a given closed loop $c$, the sign structure is now written as
\begin{eqnarray}
  \label{eq:sign}
  \tau_{c}=\tau_{\text{ps}}\times\tau_{h}
\end{eqnarray}
with the phase string signs in $\tau_{\text{ps}}=(-1)^{N_{h}^{\downarrow }[c]}$ and the fermionic signs in $\tau_{h}=(-1)^{N_{\text{ex}}^{h}[c]}$.  Here $N_{h}^{\downarrow }[c]$ is the total number of exchanges between the dopant holes and down spins and $N_{\text{ex}}^{h}[c]$ denotes the total number of exchanges among dopant holes. The phase string signs in $\tau_{\text{ps}}$ are the dynamical frustrations turning the long range antiferromagnetic state into short range spin liquid state.\cite{Weng2007} Unlike the full fermionic signs in the Fermi liquid state, the fermionic statistics in $\tau_{h}$ is only activated for the dopant holes which has the particle number $p$ instead the total number of the electrons $n_{e}=1-p$. The Luttinger's theorem is now modified as
\begin{eqnarray}
  \label{eq:Lutt} -2{\frac{V_{F}}{(2\pi)^{2}}}=n_{e}-1\mod2
\end{eqnarray}
The minus sign on the left represents the hole pockets. This modified Luttinger theorem is also obtained by using the elegant Oshikawa argument\cite{Oshikawa2000} in Refs. \onlinecite{Senthil2003,Senthil2004}.

\subsection{The phase string theory of the LvvFL}
To capture the sign structure, the phase string theory describe LvvFL as a projected product state\cite{Weng2011}
\begin{eqnarray}
  \label{eq:ps}
  |\text{LvvFL}\rangle_{\text{ps}}=\mathcal{P}\left( |\widetilde{\text{RVB}}
  \rangle_{b}\bigotimes|\Phi\rangle_{a}\right)
\end{eqnarray}
Here the phase string signs, $\tau_{\text{ps}}$ in Eq. (\ref{eq:sign}), are embedded in a half-filled ($n=1$) bosonic resonant-valence-bond state $|\widetilde{\text{RVB}}\rangle_{b}$, described by the mutual Chern-Simons theory.\cite{Kou2005,Weng2011} The spare fermionic signs in Eq. (\ref{eq:sign}) are carried by the Fermi pockets described by $|\Phi\rangle_{a}$.

Here $|\Phi\rangle_{a}$ is naturally an LvvFL state in the PGMS when the pairing of the fermions $a_{i\sigma}$ is destroyed, say, by a strong magnetic field.\cite{Weng2011} It will lead to a residual DOS as observed in the NMR experiment. On the other hand, the neutral spin background $|\widetilde{\text{RVB}} \rangle _{b}$, which evolves from a long-range antiferromagnetic state to a short-range spin liquid state at finite doping, will be responsible for the PG behavior of the NMR Knight shift, $1/T_{1}$ and $1/T_{2}$, as have been previously shown in Ref. \onlinecite{Gu2005} (cf. Fig. 7).

\subsection{The dopon theory for the LvvFL}\label{sec:dopon}
\begin{figure}
\includegraphics[width=4.5cm]{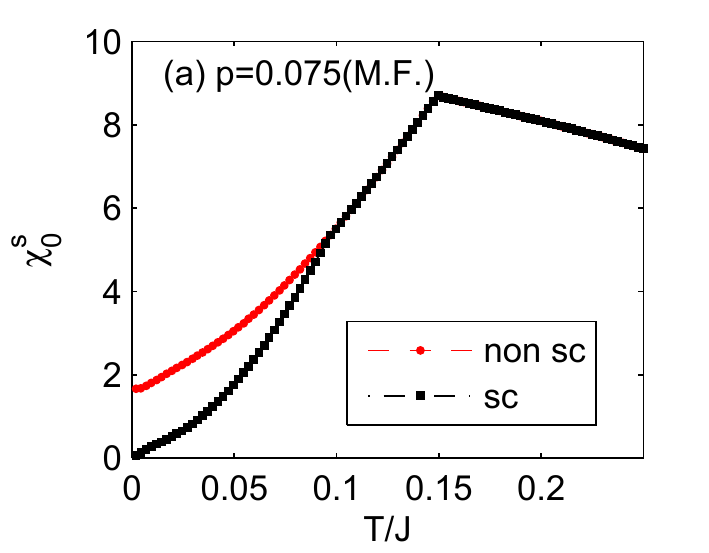}\includegraphics[width=4.5cm]{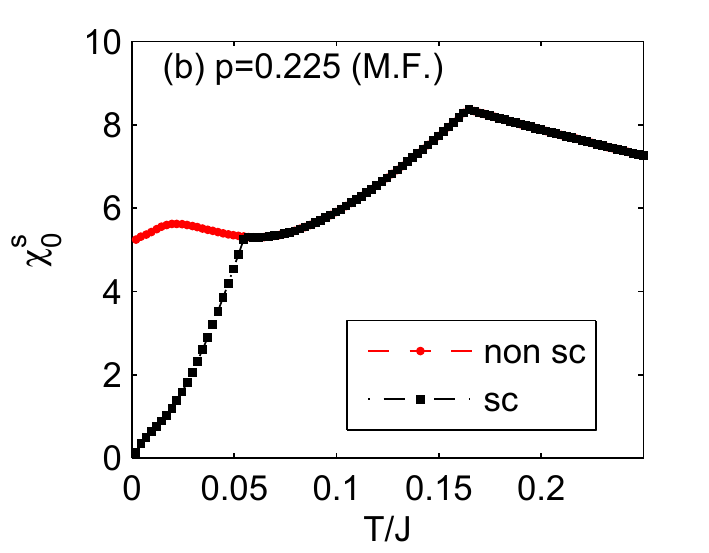}\\
\includegraphics[width=4.5cm]{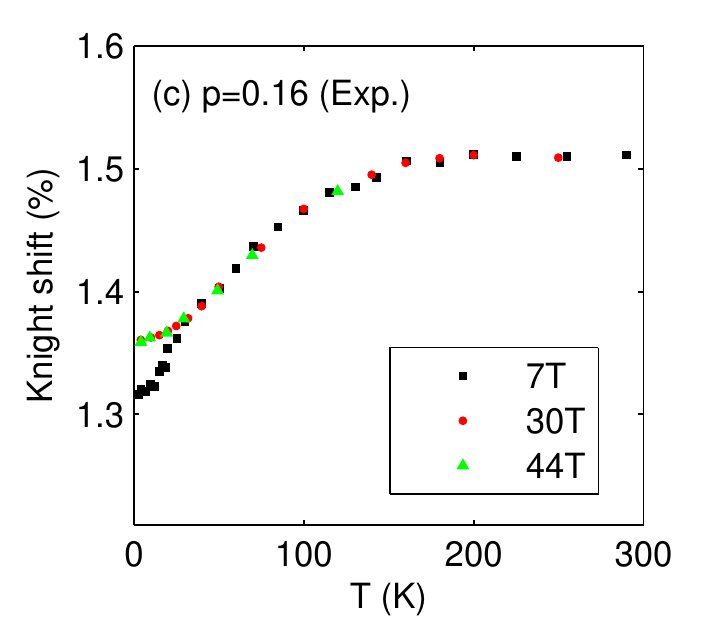}\includegraphics[width=4.5cm]{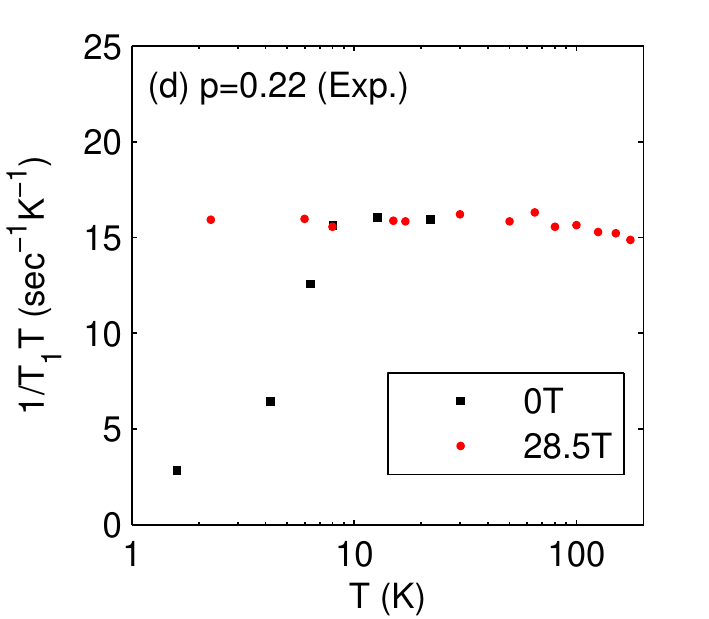}
\caption{[Color online] The mean field $T$-dependence of the uniform spin susceptibilities from the dopon theory in superconducting (SC) and non-superconducting (non SC) states, for (a) underdoped case $x=0.075$ and (b) overdoped one $x=0.225$.  (c) The experimental Knight shift for optimal La-Bi2201 under different magnetic fields; (d) The experimental $1/T_1T$  for overdoped Bi-La2201 $p=0.22$.  The experimental data under strong magnetic fields above 26T are for non-superconducting (non SC) states.
}
\label{fig:chi} 
\end{figure}
The dopon theory developed in Refs. \onlinecite{Ribeiro2005,Ribeiro2006} can also describe the LvvFL. In the dopon theory, the LvvFL is described by the following many-body trial wave function
\begin{eqnarray}
  |\text{LvvFL}\rangle_{\text{dopon}}=\mathcal{P}\left(|\text{RVB}\rangle_f\bigotimes|\Phi\rangle_d\right)
\end{eqnarray}
Here $|\text{RVB}\rangle_f$ is the half-filled ($n=1$) fermionic RVB state (spinons) and $|\Phi\rangle_d$ describes the Fermi pockets (dopons). In the dopon theory, the phase string signs are simply treated by the projected short range spin liquid. Both the spinons and dopons are fermions.

In this paper, we will give some mean field quantitative calculations. The dopon theory has the mean field Hamiltonian\cite{Ribeiro2005,Ribeiro2006}
\begin{eqnarray}
  H_{\text{MF}}&&=\begin{pmatrix}
f_{\mathbf{k}}^\dag & d_{\mathbf{k}}^\dag \end{pmatrix}\begin{pmatrix}
\alpha_{\mathbf{k}}^z\sigma_z+\alpha_k^x\sigma_x & \beta_{\mathbf{k}}\sigma_z\\
\beta_{\mathbf{k}}\sigma_z & \gamma_{\mathbf{k}}\sigma_z
\end{pmatrix}\begin{pmatrix} f_{\mathbf{k}} \\ d_{\mathbf{k}}
\end{pmatrix}\nonumber\\ +&&
{3J_{\text{eff}}\over4}N(\chi^2+\Delta^2)-N\mu_d(1-p)-2N b_0 b_1
\end{eqnarray}
where $\alpha_{\mathbf{k}}^z=-\left({3J_{\text{eff}}\over4} \chi-t_1p\right) (\cos k_x+\cos k_y)-\mu_f$, $\alpha_{\mathbf{k}}^x=-{3J_{\text{eff}}\over4}\Delta(\cos k_x-\cos k_y)$, $\gamma_{\mathbf{k}}=2t_2 \cos k_x\cos k_y+ t_3(\cos 2k_x+\cos 2k_y)-\mu_d$ and $\beta_{\mathbf{k}}={3b_0\over8}[t_1(\cos k_x+\cos k_y)+2t_2\cos k_x\cos k_y +t_3(\cos 2k_x+\cos 2k_y)]+b_1$.  $f_{\mathbf{k}}^\dag=\begin{pmatrix} f_{\mathbf{k}\uparrow}^\dag & f_{-\mathbf{k}\downarrow} \end{pmatrix}$, $d_{\mathbf{k}}^\dag=\begin{pmatrix} d_{\mathbf{k}\uparrow}^\dag & d_{-\mathbf{k}\downarrow} \end{pmatrix}$ and $\sigma_z$, $\sigma_x$ are the Pauli matrices. The mean field Hamiltonian has the eigenvalues $\pm\epsilon_{\mathbf{k}1}$ and $\pm\epsilon_{\mathbf{k}2}$ with the definitions 
\begin{eqnarray}
  \epsilon_{\mathbf{k}1}=\sqrt{\rho_{\mathbf{k}}-\sqrt{\tau_\mathbf{k}}},\quad \epsilon_{\mathbf{k}2}=\sqrt{\rho_\mathbf{k}+\sqrt{\tau_\mathbf{k}}}.
  \label{eq:epsilon12}
\end{eqnarray}
Here $\rho_\mathbf{k}=\beta_\mathbf{k}^2+\frac{1}{2}[(\alpha_\mathbf{k}^z)^2+(\alpha_\mathbf{k}^x)^2+\gamma_\mathbf{k}^2]$ and $\tau_\mathbf{k}=\beta_\mathbf{k}^2[(\alpha_\mathbf{k}^z+\gamma_\mathbf{k})^2+(\alpha_\mathbf{k}^x)^2]+\frac{1}{4}[(\alpha_\mathbf{k}^z)^2+(\alpha_\mathbf{k}^x)^2-\gamma_\mathbf{k}^2]^2$. Therefore, the mean field free energy is given as
\begin{eqnarray}
  F_\text{MF}&=&-\frac{1}{\beta}\sum_\mathbf{k}\ln[(1+\cosh(\beta\epsilon_{\mathbf{k}1}))(1+\cosh(\beta\epsilon_{\mathbf{k}2}))]\nonumber\\
  &+&\frac{3J_\text{eff}}{4}N(\chi^2+\Delta^2)-N\mu_d(1-x)-2Nb_0b_1
  \label{eq:fmf}
\end{eqnarray} 
To obtain the proper phase diagram, the parameters in the dopon theory are
tuned as follows: $J_{\text{eff}}=(1-p)^2J$, $t_1=1.5J$, $t_2={p\over0.3}J$ and
$t_3=J-{p\over0.6}J$. The order parameters $\chi$, $\Delta$, $b_0$, $b_1$,
$\mu_f$ and $\mu_d$ are determined by minimizing the free energy.\cite{Ribeiro2006} The spectral weight in ARPES and local DOS in STM/STS extracted from the electron Green's function are calculated in the previous works\cite{Ribeiro2005,Ribeiro2006a} and are not repeated here.  

There are two different normal states above $T_c$: the LvvFL in the upderdoped regime  and the Fermi liquid in the overdoped regime. When the magnetic fields drive the normal states in the T=0K limit, we simply set $b_0,b_1=0$ and $\Delta=0$ in the LvvFL and the Fermi liquid, respectively. Under magnetic fields $h$, the mean field eigenvalues are changed into $\epsilon_{\mathbf{k}(1,2)}(h)=\epsilon_{\mathbf{k}(1,2)}-h$. The mean field spin susceptibility $\chi_0^s=-\frac{\partial^2 F_\text{MF}}{\partial h^2}$ is given as
\begin{eqnarray}
  \chi_0^s=\frac{1}{2}\sum_\mathbf{k}[\beta\text{sech}^2(\beta\epsilon_{\mathbf{k}1}/2)+\beta\text{sech}^2(\beta\epsilon_{\mathbf{k}2}/2)]
  \label{eq:chi_0}
\end{eqnarray}
The mean field results of $T$-dependent uniform spin susceptibility $\chi_0^s$ for $p=0.075$ (underdoped) and $p=0.225$ (overdoped) are shown in  Fig.  \ref{fig:chi} (a) and (b). To compare with the NMR measurements, we also plot the $T$-dependent Knight shift $K_c$ for the optimal sample ($p=0.16$) and $1/T_1T$ for the overdoped sample ($p=0.22$) in Fig. \ref{fig:chi} (c) and (d). The mean field results coincide semi-quantitatively with the experimental data. There are some unphysical nonanalytic peaks in the mean field spin susceptibility around $T/J\sim 0.16$ in Fig.\ref{fig:chi} (a) and (b), not observed in the NMR data. They are able to be removed by including the gauge field fluctuations which is beyond the mean field calculation and is not our concern in this paper.   

\section{Summary}
We study the normal metallic state at zero temperature in the underdoped cuprates which we call the PGMS in this paper. Based on the NMR measurements, we identify the PGMS as an LvvFL. The LvvFL has well-defined low energy excitations but with the Fermi pockets of smaller volume violating the Luttinger's theorem in the standard Landau Fermi liquid. The LvvFL is a quantum liquid and has the translational symmetry. To understand the LvvFL, we describe the underdoped cuprates as the doped Mott insulator and employ two different theories, the phase string theory and the dopon theory, to capture the physics of the LvvFL. Both theories are in terms of two components: a background RVB state and a Fermi liquid with small Fermi pockets embedded in the short-range antiferromagnetic spin background. We mainly focus on the dopon theory and obtain some semi-quantitative results on the mean field level in a good agreement with the experimental data.

\section{acknowledgment}

J. W. Mei thanks Lu Li, Hai-Hu Wen, Yao-Min Dai, Jian-Qiao Meng and Xin-Jiang Zhou for useful discussions. J. W. Mei also thanks T. M. Rice for the suggestions on the revision of this paper. JWM and ZYW are supported by NSFC grant No. 10834003 and National Program for Basic Research of MOST Nos.  2009CB929402 and 2010CB923003. XGW is supported by NSF Grant No. DMR-1005541 and NSFC 11074140. Work in Okayama was supported in part by MEXT Grant No.  22103004. Work in IOP was supported by CAS and NSFC. A portion of this work was performed at the National High Magnetic Field Laboratory, which is supported by NSF, the State of Florida, and DOE . SK and GQZ thank A.P. Reyes and P.L. Kuhns for help.
\bibliography{LvvFL0}

\end{document}